\documentclass[aps,pre,twocolumn,superscriptaddress,showpacs]{revtex4}

\usepackage{amsmath}
\usepackage{amssymb}
\usepackage{graphicx}

\newcommand{\sgn}{\mathop{\mathrm{sgn}}}
\newcommand{\ie}{i.e.\ }

\begin{document}
%
%
\title{Wave-driven dynamo action in spherical MHD systems}
\author{K.~Reuter}
\affiliation{Max-Planck-Institut f{\"u}r Plasmaphysik, EURATOM Association,
Boltzmannstra{\ss}e 2, D-85748 Garching, Germany}
\author{F.~Jenko}
\affiliation{Max-Planck-Institut f{\"u}r Plasmaphysik, EURATOM Association,
Boltzmannstra{\ss}e 2, D-85748 Garching, Germany}
\author{A.~Tilgner}
\affiliation{Institute of Geophysics, University of G{\"o}ttingen,
Friedrich-Hund-Platz 1, 37077 G{\"o}ttingen, Germany}
\author{C.~B.~Forest}
\affiliation{Department of Physics, University of Wisconsin-Madison, 1150
University Avenue, Madison, Wisconsin 53706, USA}
\date{\today}

\begin{abstract}
Hydrodynamic and magnetohydrodynamic numerical studies of a mechanically forced
two-vortex flow inside a sphere are reported.  The simulations are performed in
the intermediate regime between the laminar flow and developed turbulence where
a hydrodynamic instability is found to generate internal waves with a
characteristic $m=2$ zonal wave number.  It is shown that this time-periodic
flow acts as a dynamo although snapshots of the flow as well as the mean flow
are not dynamos.  The magnetic fields' growth rate exhibits resonance effects
depending on the wave frequency.  Furthermore, a cyclic self-killing and
self-recovering dynamo based on the relative alignment of the velocity and
magnetic fields is presented.  The phenomena are explained in terms of a mixing
of non-orthogonal eigenstates of the time dependent linear operator of the
magnetic induction equation.  The potential relevance of this mechanism to
dynamo experiments is discussed.
\end{abstract}


\maketitle

\section{Introduction}
It is well-established theoretically that magnetic fields which emanate from the
Earth and from celestial bodies are generated by flows of electrically
conducting fluids in their interiors, a mechanism which is commonly referred to
as the magnetohydrodynamic (MHD) dynamo effect \cite{Moffatt, MagneticUniverse}.
In recent years, significant effort was devoted to the experimental verification
of dynamo theory in simply connected spherical \cite{Nornberg2006a, Spence2006,
Spence2007} and cylindrical \cite{Monchaux2007, Ravelet2008} impeller-driven
flows of liquid sodium.  While the latter setup has ultimately demonstrated a
self-excited dynamo, it has turned out that turbulence strongly inhibits the
dynamo process in both experiments.  A review on dynamo experiments is given in
Ref.~\cite{Stefani_ZAMM_2008}.
%
%
Detrimental effects of turbulence on the excitation threshold of large-scale
dynamos were confirmed by numerical simulations of MHD systems with well-defined
mean-flows in periodic-box geometry \cite{Ponty2005, Minnini2006, Laval2006,
Ponty_NJP_2007} as well as in spherically bounded geometry \cite{Bayliss2007,
Gissinger_PRL_2008, Reuter_NJP_2008}.

In the present paper, we study numerically the regime between turbulence and the
laminar state of an electrically conducting two-vortex flow inside a sphere
which is similar to the flow realized in the Madison Dynamo Experiment
\cite{Nornberg2006a, Spence2006, Spence2007}.
The structure of the manuscript is as follows.
The numerical model is presented in section \ref{sec:model}.
Section \ref{sec:hydro} discusses purely hydrodynamic simulations which are
performed in the sub-turbulent regime.  It is found that a hydrodynamic
instability leads to traveling internal waves, their dominant component being
symmetric under rotations by $\pi$.
In section \ref{sec:MHD}, we turn to MHD investigations of the flows introduced
in the previous section.  It is found that the presence of the waves which
correspond to smoothly oscillating large-scale fluctuations supports the dynamo
instability.
Artificially changing the wave's frequency yields a resonance effect, i.e.\ the
dynamo is found to operate most efficiently at certain frequencies whereas
dynamo action ceases for frequencies which are too low or too high.
Moreover, a nonlinear dynamo is reported which undergoes a cycle of
self-killing and self-recovering events.  We find that this behavior is related
to phase shifts between characteristic magnetic and hydrodynamic oscillations
which are imposed during the transitions.  These phase shifts translate to
changes in the relative alignment of the velocity and magnetic fields.
Section \ref{sec:conclusion} concludes with a discussion and interpretation of
the numerically obtained results.  We interpret the MHD phenomena in the
framework of a dynamo mechanism based on the mixing of non-orthogonal
eigenstates of the time-dependent linear operator of the induction equation
\cite{Tilgner_PRL_2008}, thus confirming its relevance to a self-consistent
dynamo model.

\section{Numerical model}
\label{sec:model}
The magnetic ($\mathbf B$) and velocity ($\mathbf v$) fields which describe an
incompressible electrically conducting fluid are governed by the induction
equation (\ref{ind_eq}) and the Navier-Stokes equation (\ref{ns_eq}),
\begin{align}
\frac{\partial \mathbf B}{\partial t} &=
\nabla \times ( \mathbf v \times \mathbf B ) +
\lambda \nabla^2 \mathbf B ,    \label{ind_eq}
\\ 
\frac{\partial \mathbf v}{\partial t} + \left(
\mathbf v \cdot \nabla \right) \mathbf v &= - \nabla p + \nu \nabla^2 \mathbf v
+ \mathbf j \times \mathbf B + \mathbf F ,  \label{ns_eq}
\end{align}
along with the constraints $\nabla \cdot \mathbf B = \nabla \cdot \mathbf v = 0$.
Here, $\lambda$ is the magnetic diffusivity, $p$ is the pressure, $\mathbf j =
\mu_0^{-1} \nabla \times \mathbf B$ is the current density, $\nu$ is the
viscosity, and $\mathbf F$ is a forcing term.
A constant mass density $\rho=1$ is assumed.  In nondimensional form, the
problem is characterized by two control parameters, the Reynolds number
$\mathrm{Re}=LV \nu^{-1}$ and the magnetic Reynolds number $\mathrm{Rm}=LV
\lambda^{-1}$, where $L$ and $V$ denote length and velocity scales which are
characteristic of the system under consideration.
We choose $L=1$ and $V=\overline{v_{\mathrm{rms}}}$, where $v_{\mathrm{rms}} =
\sqrt{\langle \mathbf{v}^2 \rangle}$ is the spatial rms velocity.  Here, the
angle brackets denote averaging in space, and the overline denotes averaging in
time which is performed during the quasi-stationary phase of the flow after all
initial bifurcations.
The characteristic timescale of the flow is given by the eddy turnover time
$\tau_{\nu}=L/V$, whereas the magnetic diffusion time $\tau_{\sigma} =
\mathrm{Rm} \, \tau_{\nu}$ is the timescale relevant to the magnetic field.

We solve numerically the MHD equations (\ref{ind_eq})-(\ref{ns_eq}) in a
sphere, using the parallel version of the \textsc{Dynamo} code
\cite{Bayliss2007,Reuter_CPC_2008}.  It employs the standard pseudo-spectral
method based on a poloidal-toroidal decomposition of the vector fields in
combination with spherical harmonic expansions.  Outer boundary conditions are
the potential field solution for the magnetic field, and, unless stated
otherwise, the zero-slip condition for the velocity field.  The forcing is
designed to produce an $s2t2$ type of flow \cite{DudleyJames1989}, which consists of
two in the toroidal direction counter-rotating hemispherical cells with
poloidal circulation in each cell, directed outwards to the poles and inwards
in the equatorial plane.  It is given by an axisymmetric localized body force
$\mathbf F$ kept constant in time, which reads in a cylindrical coordinate
system ($s,\phi,z$)
\begin{align}
    F_{s} &=0, \notag            \\
    F_{\phi} &= \epsilon \sgn(z) s^3 r_d^{-3} \sin{\frac{\pi s}{2 r_d}} +
    \gamma,   \label{eq:forcing}    \\
    F_{z}    &= (1-\epsilon) \sgn(z) \sin{\frac{\pi s}{r_d}} + \delta. \notag
\end{align}
Assuming a sphere of radius $r=1$, the driving is applied within
$0.25 < |z| < 0.55$, $s < r_d = 0.29$, with the parameters kept constant at
$\epsilon=0.1$, $\gamma=0.05$, $\delta=0.3$.
For sufficiently large $\mathrm{Rm}$, the resulting flow is a dynamo, the
($\ell=m=1$) magnetic eigenmode growing fastest.  The generation of this
transverse dipole can be understood in the axisymmetric flow using a simple
frozen-flux picture based on stretching, twisting, and folding of magnetic field
lines \cite{Nornberg2006b}.

In all simulations, the spherical harmonic expansions are triangularly truncated
at degree and order $\ell_{\mathrm{max}}=m_{\mathrm{max}}=20$, while $160$
uniformly spaced radial points are used.  To verify the simulation results
obtained at the aforementioned resolution we have repeated a nonlinear run
(the computation of flow C which is introduced in the following section) using
$\ell_{\mathrm{max}}=m_{\mathrm{max}}=30$ and 320 gridpoints in radial
direction.  The bifurcations towards flow C as well as the growthrate of the
magnetic field in the MHD case were fully reproduced which indicates that
converged solutions are obtained already at the coarser resolution.  Moreover,
power spectra of the velocity fields in terms of $\ell$ show a drop-off by three
to four orders of magnitude between the dominant and the highest wavenumber.
The magnetic field is resolved similarly well.  Relevant spectra are shown and
discussed in the following sections, cf.\ Figs.~\ref{fig04} and \ref{fig12}.
%
%
After having introduced the numerical setup we now turn to an investigation of
the hydrodynamics of the flow before we consider full MHD cases.

\section{Hydrodynamic studies}
\label{sec:hydro}
We first investigate the dynamics of the unmagnetized flow, \ie
Eq.~(\ref{ns_eq}) is integrated forward in time with $\mathbf B = \mathbf 0$.
Starting from a fluid at rest, momentum is injected by the body force $\mathbf
F$, and the time integration is performed until a statistically stationary state
is reached.  This procedure is repeated to scan over various decreasing
viscosities, thus increasing $\mathrm{Re}$.
For $\mathrm{Re}<65$, the flow reaches a stationary axisymmetric ($m=0$) state.  Above
this threshold, the flow becomes hydrodynamically unstable.
\begin{figure}   \includegraphics[width=\columnwidth]{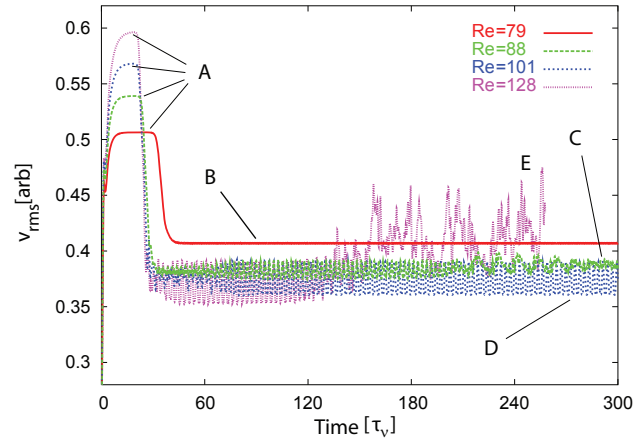}
\caption{\label{fig01}Time traces of $v_{\mathrm{rms}}$ for four $s2t2$ flows
driven by the body force $\mathbf F$, at $\mathrm{Re}$ being moderately above the stable
region $\mathrm{Re}<65$, illustrating bifurcation sequences.
}
\end{figure}
To exemplify subsequent temporal transitions, Fig.~\ref{fig01} shows time traces
of $v_{\mathrm{rms}}$ at $\mathrm{Re}=79, 88, 101, 128$, the quasi-stationary
states being labeled B, C, D, E.  In each case, the flow first reaches an
axisymmetric configuration which is characterized by an initial plateau (A).
Simultaneously, non-axisymmetric modes with even $\ell$ and zonal wave number
$m=2$ grow fastest at an exponential rate.  Shortly before saturation occurs,
the symmetry breaking becomes visible in physical space near the polar regions
of the sphere as indicated in Fig.~\ref{fig02}~(a).  The previously circular
cross sections of the jets towards the poles are stretched elliptically as
it is depicted in Fig.~\ref{fig02}~(b-d) for the flow D.
\begin{figure}   \centering
\includegraphics[width=\columnwidth]{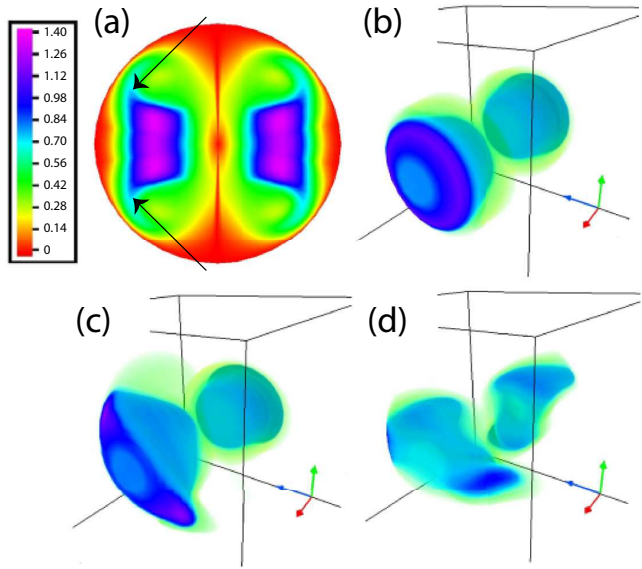}
\caption{\label{fig02}
Illustrations of the symmetry breaking in the flow towards state D.
(a)    Poloidal cross section of the energy density $\propto\mathbf{v}^2$ during
the axisymmetric phase A.  Arrows indicate the region where
elliptical stretching occurs.
(b-d)  Time series depicting the spatial structure of large amplitudes of
$\mathbf{v}^2$ inside the sphere $r<0.75$ as the instability grows.
}
\end{figure}
The nonlinear saturation of this instability manifests itself in a relaxation of
$v_{\mathrm{rms}}$, cf. Fig.~\ref{fig01}.
In physical space, traveling waves with $m=2$ symmetry emerge, their spatial
structure being illustrated in Fig.~\ref{fig03}.
\begin{figure}   \centering
\includegraphics[width=\columnwidth]{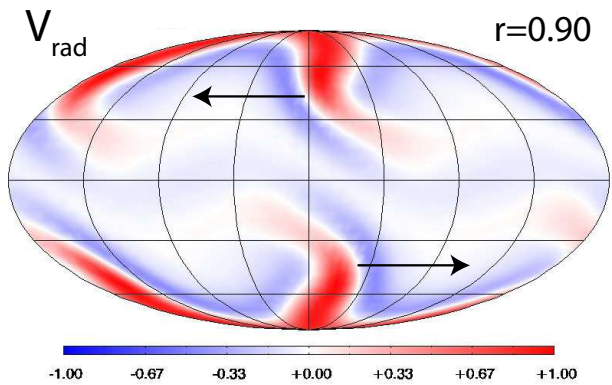}
\caption{\label{fig03}
Mollweide projection of the radial component of a velocity field snapshot at
90\% of the sphere's radius during the statistically stationary state D.  Arrows
indicate the direction of wave propagation.
}
\end{figure}
The wave propagation is directed opposite to the sense of rotation of the
mean-flow in the individual hemisphere as prescribed by the driving, with the
wave vector pointing in the zonal direction.
With increasing $\mathrm{Re}$, the latitudinal extent of the wave grows towards
the equator, coupling the two hemispheres and causing oscillatory fluid motions
(cf. Fig.~\ref{fig01}, states C and D compared to state B).
As a result, the wave frequency decreases with increasing $\mathrm{Re}$.
With $\mathrm{Rm} \approx 44$ (see below), the frequencies in the lab frame turn
out to be fast on the resistive timescale, namely
$f_{\mathrm{C}} \approx 19.08 \tau_{\sigma}^{-1} = 0.43 \tau_{\nu}^{-1}$ and
$f_{\mathrm{D}} \approx 16.71 \tau_{\sigma}^{-1} = 0.39 \tau_{\nu}^{-1}$.
For $\mathrm{Re} > 125$, turbulence develops, and the $m=2$ wave feature loses
its clear structure (state E).

\begin{figure}   \centering
\includegraphics[width=\columnwidth]{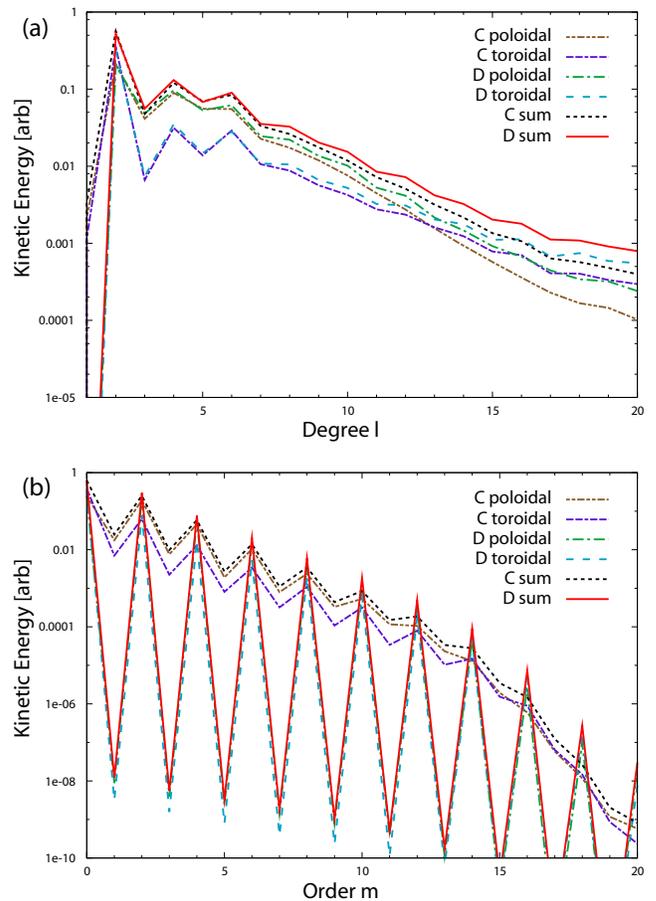}
\caption{\label{fig04}
Time-averaged spectra of the kinetic energy density in the flows C and D as
functions of spherical harmonic degree $\ell$ (a) and order $m$ (b).}
\end{figure}

Fig.~\ref{fig04} displays time averaged energy spectra of the flows C and D in
terms of spherical harmonic degree $\ell$ and order $m$.
The time averaging was performed for several hundred $\tau_\nu$ to assure
convergence.
The $\ell$-spectra in Fig.~\ref{fig04}~(a) peak at $\ell=2$ for both flows which
is due to the imposed forcing.  The spectra are overall similar except for the
strength of the $l=1$ modes which refer to global circulation, however, these
modes are dynamically not important in flow C and even less in D due to their
small amplitudes.  The spectra of flow C drop off somewhat faster than those of
flow D which is to be expected due to the larger value of the kinematic
viscosity in run C.
The spectra in terms of the order $m$ shown in Fig.~\ref{fig04}~(b), however, yield
differences between C and D concerning the amplitude of the modes with odd
wavenumbers.  In both cases, the spectra peak at $m=0$ which is due to the
axisymmetric background flow.  The dominant peak connected with non-axisymmetric
modes is found at $m=2$ and caused by the wave feature, followed by peaks at
higher harmonics with even $m$.  The origin of the preference for even modes in
flow D remains unclear.  During the initial bifurcation sequence, the amplitudes
of the odd $m$-modes drop off suddenly around $\tau_\nu \approx 70$ after a
phase spectrally similar to C before the statistically stationary state D
($\mathrm{Re}=101$) is reached.  The corresponding change of the flow becomes
apparent in Fig.~\ref{fig01} as a change in the oscillatory behaviour of the
respective time trace.  Conversely, the flow C at $\mathrm{Re}=88$ starts off with
odd $m$-modes at small amplitude after the initial transition A.  The
respective time trace in Fig.~\ref{fig01} shows regular oscillations until,
at $\tau_\nu \approx 210$, odd $m$-modes have grown sufficiently strong to
become dynamically important.  The change in the flow towards the statistically
stationary state C manifests itself in a more irregular time trace of
$v_{\mathrm{rms}}$.
Flows with metastable states switching between the characteristics of flow C and
D were not observed, although this possibility may exist.
It should be noted that, when going to higher $\mathrm{Re}$, the $m$-spectra
tend to be smoothed.  A local peak at $m=2$ is still found, and can be
interpreted as the signature of the most unstable modes excited by the
hydrodynamic instability.

In the case of flow D, the temporal evolution of the dominant components of the
spectrum is shown in the upper panel of Fig.~\ref{fig09}.
The time trace of the ($\ell=2,m=0$) axisymmetric background flow [which will be
abbreviated (2,0) in the following] is modulated sinusoidally at frequency $f_D$
via nonlinear interactions with other modes.
The $m=2$ wave feature corresponds to modes oscillating at a time-periodic
non-sinusoidal pattern at an amplitude which is about $10\%$ of the amplitude of
the (2,0) background flow. 
In the case of flow C, the picture is overall similar, although the time traces
show different characteristics concerning their periodicity and shape.

Additional numerical experiments were performed to better understand the origin
of the wave phenomenon.
Outer shear-layers which exist due to the zero-slip condition on the velocity
field prove to be crucial for the wave generation.  This boundary condition
requires the three vector components of the velocity field to be zero at the
outer boundary.  It is intended to be an approximation of the experimental
reality where liquid sodium is confined by a solid wall.
Applying the stress-free boundary condition instead in a numerical experiment
while leaving the forcing $\mathbf{F}$ unchanged fundamentally alters the nature
of the hydrodynamic instability.  The stress free boundary condition requires
the radial component of the velocity field to be zero at the boundary, whereas
the angular components evolve under the constraint of zero stress.  As a
result, a flow pattern with a strong zonal flow next to the boundary develops in
each hemisphere.  Stationary states with waves comparable to the flows C and D
are not observed.
These findings support the potential relevance of the results presented in this
paper to existing dynamo experiments where boundary layers with strong shear
exist in the flows next to the outer walls.

Moreover, the fact that $s2t2$ flows were studied in the past by several groups,
each of them using slightly different forcing functions or prescribed flows,
motivates us to investigate the role of the forcing term on the wave generation.
Applying the zero-slip condition, wavy states similar to C and D are found in
the sub-turbulent regime when the flow is driven by the localized body force
given in \cite{Bayliss2007,Gissinger_PRL_2008}, and, similarly, when the global
$s2t2$ profiles discussed in \cite{DudleyJames1989} are employed to drive the
flow.
%
%
%
In addition to the axisymmetric component, non-axisymmetric $m=2$ modes are found
to dominate in these flows, however, modes with $m=1$ and $m=3$ wave numbers are
dynamically important as well.
A main difference between the three forcing schemes consists in the distance
between the active region where momentum is injected, and the outer boundary.
This distance is largest in our impeller model, Eq.~\ref{eq:forcing}, and is
chosen to be smaller in the latter forcing schemes.  The decrease affects the
thickness of the boundary layer and thereby the rate of strain which in turn may
govern which zonal mode $m$ is most unstable.
A final comment on the influence of the forcing is made on the structure of the
magnetic field in case of a dynamo.  It turns out that a flow forced by
Eq.~\ref{eq:forcing} favors an $m=1$ dipole mode in the laminar case, in the
intermediate sub-turbulent range on which we focus in the present paper, as well
as in the presence of developed turbulence.  An $m=0$ axial dipole as it was
reported in \cite{Bayliss2007,Gissinger_PRL_2008} has not been found using
Eq.~\ref{eq:forcing}.  We have, however, numerically confirmed its existence in
turbulent flows when the forcing reported in \cite{Bayliss2007,
Gissinger_PRL_2008} is applied.

A more detailed investigation of the hydrodynamics of the $s2t2$ flow is
certainly required to shed light on the plausible conjecture that the previously
discussed hydrodynamic instability and wave formation are universal features of
the spherically bounded $s2t2$ flow.  The corresponding study, however, goes
beyond the scope of the present manuscript.  We now proceed towards the actual
goal of the paper, namely the investigation of the dynamo properties of the
sub-turbulent flows C and D.

\section{MHD investigations}
\label{sec:MHD}
The $s2t2$ class of dynamos has been the focus of several numerical
investigations, pioneered by the work of Dudley and James \cite{DudleyJames1989}
who solved the induction equation for a stationary prescribed two-vortex flow.
Later---mainly concentrating on the role of turbulence on the dynamo
process---fully time dependent flows in nonlinear MHD simulations were studied
\cite{Bayliss2007,Gissinger_PRL_2008,Reuter_NJP_2008}.
Until now, however, the effect of smoothly fluctuating fluid motions below the
transition to developed turbulence has not been considered explicitly.  This
motivates us to investigate the temporal evolution of seed magnetic fields in
the flows C and D.  To this end, the magnetic field $\mathbf B$ is initialized
with pseudo-random noise at small amplitude, and the induction equation is
solved numerically using four different types of velocity fields $\mathbf v$
based on C and D:
%
%
%
%
%
(i) time dependent flows $\mathbf v(\mathbf x, t)$, \ie the coupled
Eqs.~(\ref{ind_eq})-(\ref{ns_eq}) are integrated;
(ii) snapshots $\mathbf v(\mathbf x, t_i),~i=1 \ldots n$ of the flows kept fixed
in time;
(iii) time averaged velocity fields $\overline{\mathbf v} = \frac{1}{T}
\int_{t_0}^{t_0+T} \mathbf v(\mathbf x, t) dt$, with $T>10^2 \tau_{\nu}$;
(iv) consecutive snapshots of the flows iterated at variable rates which allows
to examine the effect of the wave frequency on the magnetic field's growth rate.
Practically, the runs of type (iv) are performed by subsequently reading
previously saved snapshots from hard disk, while solving Eq.~(\ref{ind_eq}).
Unless stated otherwise, all MHD runs discussed in the following are performed
at $\mathrm{Rm} \approx 44$.

\begin{figure}   \includegraphics[width=\columnwidth]{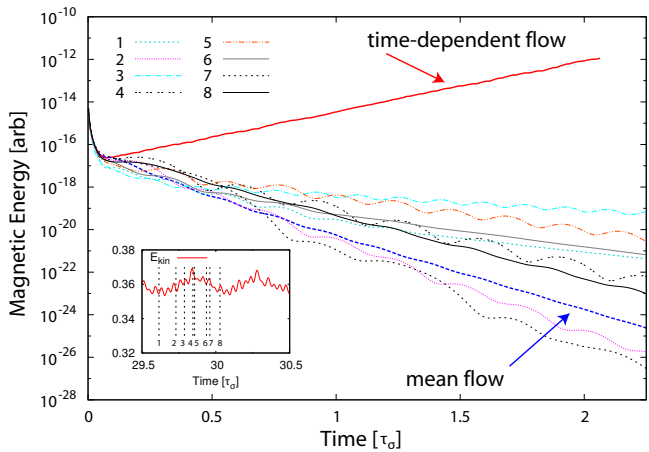}
\caption{\label{fig05}Energies of the magnetic field over time, from simulations
(i)-(iii) based on the flow C ($\mathrm{Re}=88$, $\mathrm{Rm}=44$).}
\end{figure}

\begin{figure}   \includegraphics[width=\columnwidth]{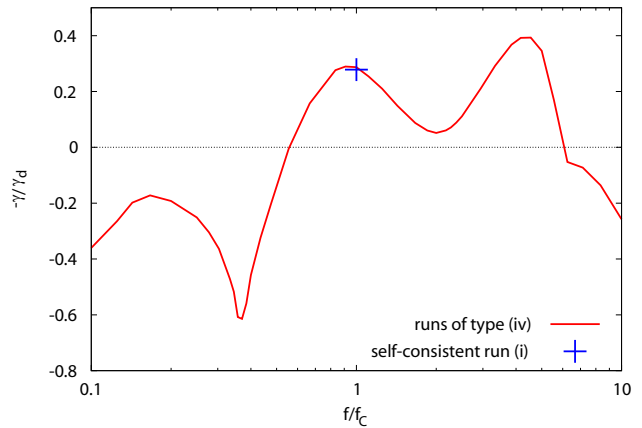}
\caption{\label{fig06}Growth rate of the magnetic field, normalized by the
negative free dipole decay rate, as a function of the relative wave frequency
$f/f_\mathrm{C}$.  The simulations were performed based on flow C at
$\mathrm{Rm}=44$.}
\end{figure}

We first consider the flow C for which Fig.~\ref{fig05} summarizes the results
of the numerical experiments (i)-(iii).
Eight randomly selected snapshots---as indicated by the inset---show decaying
magnetic fields.  Moreover, the mean-flow is not a dynamo.  Only the time
dependent flow (i) shows a growing magnetic field solution.  Clearly, the
presence of time dependence in the velocity field in the form of periodic wave
motion causes the magnetic field to grow.
In fact, in the regime around $\mathrm{Re} \approx 100$, the dynamo
threshold of the mean flow is $\mathrm{Rm}_\mathrm{c}(\overline{\mathbf v})
\approx 55$, which is significantly larger than the threshold of the time
dependent flows, being $\mathrm{Rm}_\mathrm{c} \approx 32$.
Fig.~\ref{fig06} displays the growth rate as a function of the relative wave
frequency $f/f_\mathrm{C}$ obtained by performing experiment (iv).  For
comparison, the growth rate calculated from the self-consistent run (i) is
included.
A window with positive growth rates ($f/f_\mathrm{C} \approx 0.55 \ldots 6.0$)
exists which contains two local maxima at $f/f_\mathrm{C} \approx 0.9$ and
$f/f_\mathrm{C} \approx 4.4$ being symptomatic of resonant behavior.  The first
maximum happens to be in good agreement with the frequency of the hydrodynamic
waves in run (i).  Negative growth rates occur when the wave is either too fast
or too slow.  The latter finding reflects the fact that snapshots of the flow
are not dynamos.
Iterating through the velocity field snapshots in reversed order, i.e.\
inverting the direction of wave propagation without changing the shape of the
waves, causes seed magnetic fields to decay in all investigated cases in the
range $f/f_\mathrm{C} = -10.0 \ldots -0.1$.
The observation of magnetic field amplification in a spherically bounded MHD
dynamo model induced by hydrodynamic waves represents the key result of our
work.
A power spectrum of the growing magnetic field obtained from the nonlinear MHD
simulation of flow C is given in Fig.~\ref{fig12}.  The magnetic field is
clearly dominated by the (1,1) dipole mode.
The dynamo based on experiment (i) saturates to a stationary state without any
time dependence in the velocity and magnetic fields.  The kinetic and magnetic
energies are roughly in equipartition.
Before we turn towards a discussion of the physics governing the magnetic field
growth in flow C, we perform an analogous investigation of the flow D.

\begin{figure}
\includegraphics[width=\columnwidth]{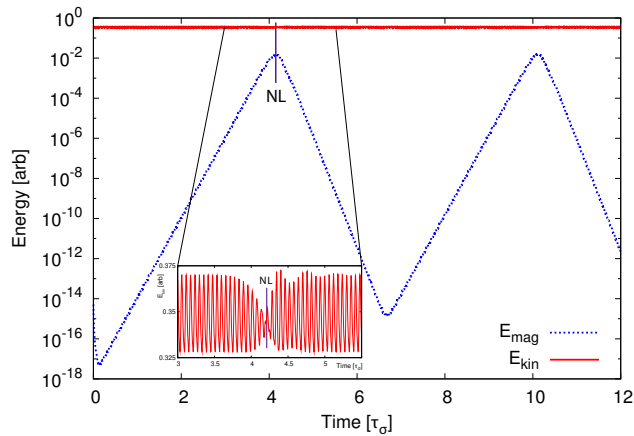}
\caption{\label{fig07}Time traces of kinetic and magnetic energy, the flow
initially being in state D ($\mathrm{Rm}=44$).  The inset shows the kinetic
energy during the transition from growth to decay.}
\end{figure}

An intriguing phenomenon is found when the evolution of magnetic fields in flow
D is studied.  Performing run (i), a seed magnetic field first grows
exponentially at a normalized rate $\gamma_{\mathrm{D}} \approx 0.46$, as it is
indicated by the time trace of the magnetic energy in Fig.~\ref{fig07}.
Nonlinear feedback via the Lorentz force acts starting from $\tau_{\sigma}
\approx 4.0$ for a period of $\sim 0.5 \tau_{\sigma}$ during which the energy
ratio peaks at $E_{mag}/E_{kin} \approx 0.04$ (labeled ``NL'' in
Fig.~\ref{fig07}).
A transition occurs to a metastable state $\mathrm{D'}$ which has a lifetime of
about $2\tau_{\sigma} \approx 90 \tau_{\nu}$.
The MHD system no longer sustains the dynamo process, as the exponential decay
of the magnetic field at rate $\gamma_{\mathrm{D'}} \approx -0.65$ indicates,
until it recovers back to the initial state D which again allows for magnetic
field growth at $\gamma_{\mathrm{D}}$, thus closing the cycle.
In the following and unless stated otherwise, the labels D and $\mathrm{D'}$
refer to the growth and decay phases as indicated in Fig.~\ref{fig07}.

To investigate the nature of the cyclic behaviour, we have repeated the
numerical experiments (ii)-(iv) using flows during both the growth phase D and
the decay phase $\mathrm{D'}$.  The mean flows and field snapshots were obtained
from run (i) at times sufficiently far away from the transitions and while the
magnetic field was negligibly small.
Experiment (ii) shows growing and decaying solutions for the magnetic field in
both cases, D and $\mathrm{D'}$.  This result does not disagree with the result
obtained from flow C at $\mathrm{Rm}=44$ since a similar situation can easily be
constructed using D by decreasing $\mathrm{Rm}$ below $44$, i.e.\ all selected
snapshots in experiment (ii) are not dynamos whereas a random seed magnetic
field grows in time.
The mean-flows during phase D and $\mathrm{D'}$ are identical by visual
inspection, which explains the finding that experiment (iii) yields identical
decay rates $\gamma_{\mathrm{DM}} \approx -0.31$.
Note that during the phase $\mathrm{D'}$ in run (i), the magnetic field decays
twice as fast as it does in the mean flow.
Finally, experiment (iv) yields resonance behaviour qualitatively similar to flow C.
To shed more light on the nature of the transitions, further investigations are
presented in the following.  The main question to be adressed is how the
velocity field, the magnetic field, or the combination of both is altered during
the transitions.

``Self-killing'' (though---to our knowledge---no ``self-recovering'') nonlinear
dynamos have been reported previously \cite{Brummell1998,Fuchs1999}.  Their
mechanism is based on a modification of the flow due to the action of the
Lorentz force away from an initial state which has supported magnetic field
growth towards a second solution of the nonlinear
Navier-Stokes-Eq.~(\ref{ns_eq}) which prevents magnetic field growth.
During the transition from D to $\mathrm{D'}$ additional exponentially decaying
smooth fluid motions on a slow timescale are excited, as the inset in
Fig.~\ref{fig07} indicates.
However, the magnetic field continues to decay at constant rate
$\gamma_{\mathrm{D'}}$ while the fluid perturbation is damped to very small
amplitude, indicating that this perturbation might be unimportant.  This
hypothesis will be confirmed by the investigations presented below.

To scrutinize the transition, we have performed the following numerical
experiments (a)-(c) using the solutions $\mathbf{v}_{\mathrm{NL}}$ and/or
$\mathbf{B}_{\mathrm{NL}}$ of the system at time $\tau_{\sigma} \approx 4.2$ as
initial conditions in the MHD equations.  The field $\mathbf B_{\mathrm{NL}}$
was rescaled to infinitesimal amplitude, comparable to the seed fields used in
previous runs.  Hence, the structure of the magnetic field is preserved whereas
an immediate backreaction on the flow is avoided.  The experiments and results
are as follows.
%
%
%
%
%
(a) Integration of the MHD equations using $\mathbf v_{\mathrm{NL}}$ and the
rescaled $\mathbf B_{\mathrm{NL}}$ as initial conditions exhibits the decay
phase at rate $\gamma_{\mathrm{D'}}$ [which is of the same duration as in (i)],
followed by the transition to growth.
(b) Integration starting from $\mathbf v_{\mathrm{NL}}$ and an infinitesimal
pseudo-random magnetic field yields magnetic field growth at rate
$\gamma_{\mathrm{D}}$.
(c) Integration starting from a velocity field obtained during $\mathrm{D'}$
different from $\mathbf v_{\mathrm{NL}}$ and from the rescaled $\mathbf
B_{\mathrm{NL}}$ (\ie introducing a phase shift between the self-consistently
adjusted configurations of the fields) leads to the same result as in run (b).

\begin{figure}   \centering
\includegraphics[width=\columnwidth]{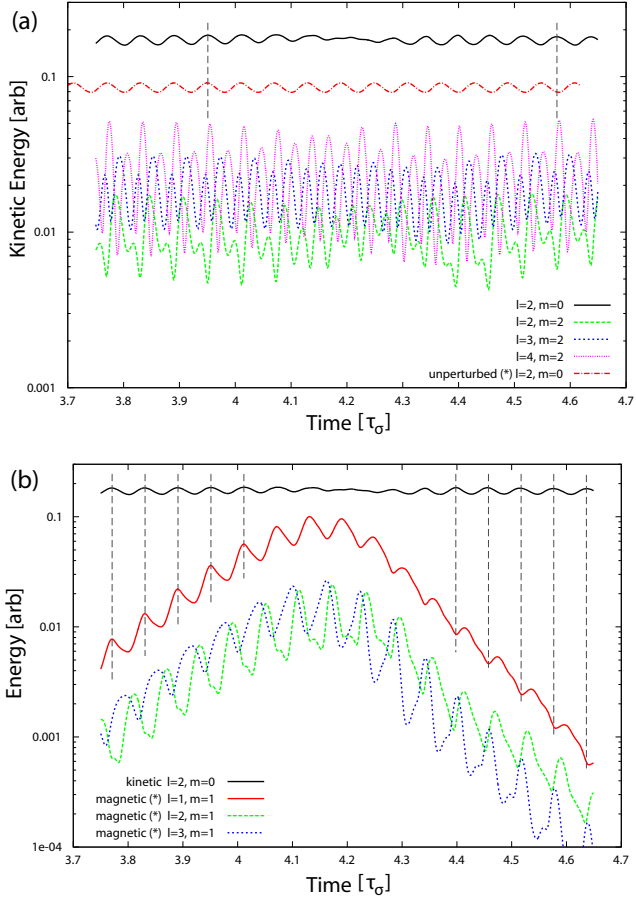}
\caption{\label{fig08}
(a)  Volume-integrated spectral energies as functions of time of the 4 dominant
modes in flow D during the transition from D to $\mathrm{D'}$ in the full MHD
simulation, c.f.\ Fig.~\ref{fig07}.  For comparison, the time trace of the (2,0)
mode of a magnetically unperturbed, purely hydrodynamic simulation of the same
flow is included, rescaled to $50\%$ of its amplitude.  The scaling is indicated
by an asterisk (*).  Vertical bars indicate the phase shift which is imposed on
the hydrodynamic oscillation.
(b)  Spectral energies as functions of time of the 3 dominant modes of the
magnetic field during the transition from D to $\mathrm{D'}$.  For comparison, the
(2,0) mode of the velocity field is displayed, cf.\ subfigure (a).  Vertical bars
illustrate a phase shift which develops between the oscillations of the magnetic
energy and the oscillations of the kinetic energy.  The magnetic energy (*) was
scaled by a factor of $10$.}
\end{figure}

\begin{figure}   \centering
\includegraphics[width=\columnwidth]{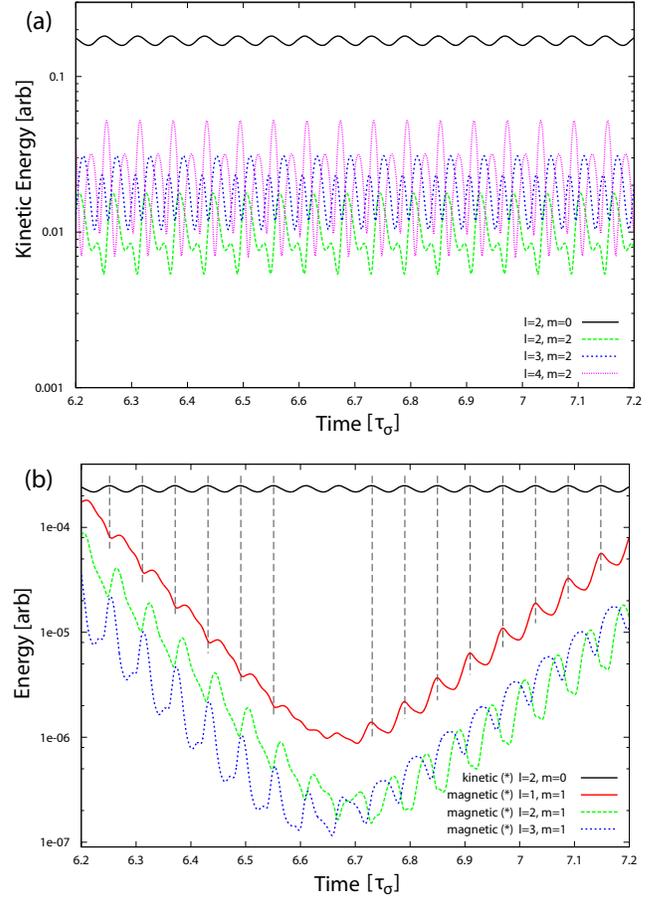}
\caption{\label{fig09}
(a)  Dominant modes of the velocity field during the transition from
$\mathrm{D'}$ to D, viz.\ Fig. \ref{fig08}~(a).
(b)  Dominant modes of the magnetic field during the transition from
$\mathrm{D'}$ to D, viz.\ Fig. \ref{fig08}~(b).  The (2,0) velocity mode from
subfigure (a) is displayed for the purpose of illustrating the phase angle
relative to the magnetic dipole mode.  For practical reasons, the amplitude of
the kinetic (magnetic) energy was scaled by a factor of $9^{-3}$ ($10^{10}$).
}
\end{figure}

We have further investigated the transitions by examining explicitly the
temporal characteristics of the spectrum.  To this end, Fig.~\ref{fig08}
displays time traces of the sum of the volume-integrated poloidal and toroidal
energies in the dominant modes during the transition from D to $\mathrm{D'}$.
As it was already pointed out in the previous section on hydrodynamics, the
(2,0) time trace of flow D is modulated sinusoidally.  Hence, it is suited to
derive the phase relative to a mode which oscillates similarly in time.
In Fig.~\ref{fig08}~(a), the dominant modes of kinetic energy obtained from the MHD
simulation are plotted over time.  These are the (2,0) mode which is responsible
for the axisymmetric background flow, and the three strongest ($m=2$) modes
which account for the bulk of the non-axisymmetric wave feature.  In addition,
the (2,0) time trace from a simulation without magnetic field is shown.  The
generating run was started using an initial condition obtained from the MHD
simulation at $\tau_\sigma = 2.0$ when the magnetic field was negligibly small.
It is seen easily by comparing the (2,0) curves that the nonlinear feedback due
to the Lorentz force causes a phase shift by $\pi$ relative to the unperturbed
flow.  Except for this phase shift, the spectral characteristics of the flow
remain unaffected.
Fig.~\ref{fig08}~(b) shows the dominant magnetic energy modes during the same
time interval, and in addition, the time trace of the (2,0) kinetic energy mode
from Fig.~\ref{fig08}~(a).  Before the transition takes place, the oscillations
in the dominant modes of the magnetic and the velocity field are in phase, as it
is indicated by vertical lines.  After the transition, the curves oscillate out
of phase by $\pi$.  In particular, the magnetic energy modes continue to
oscillate at their characteristic frequencies during the transition whereas the
kinetic energy modes experience a phase shift by $\pi$.  This phase shift is
related to a change in the relative alignment of the velocity and magnetic
fields which will be discussed in detail in the following paragraph.
The relevant spectral components during the transition from $\mathrm{D'}$ to D
are displayed in Fig.~\ref{fig09}.
It is readily seen in subfigure (a) that the velocity field modes are completely
unaffected by the change from decay to growth of the magnetic field.  The
magnetic time traces switch from out-of-phase oscillations to oscillations which
are in phase with the dominant hydrodynamic oscillations as subfigure (b)
indicates.

It is illustrative to translate the phase shifts to the respective changes in
physical space.
\begin{figure}   \centering
\includegraphics[width=\columnwidth]{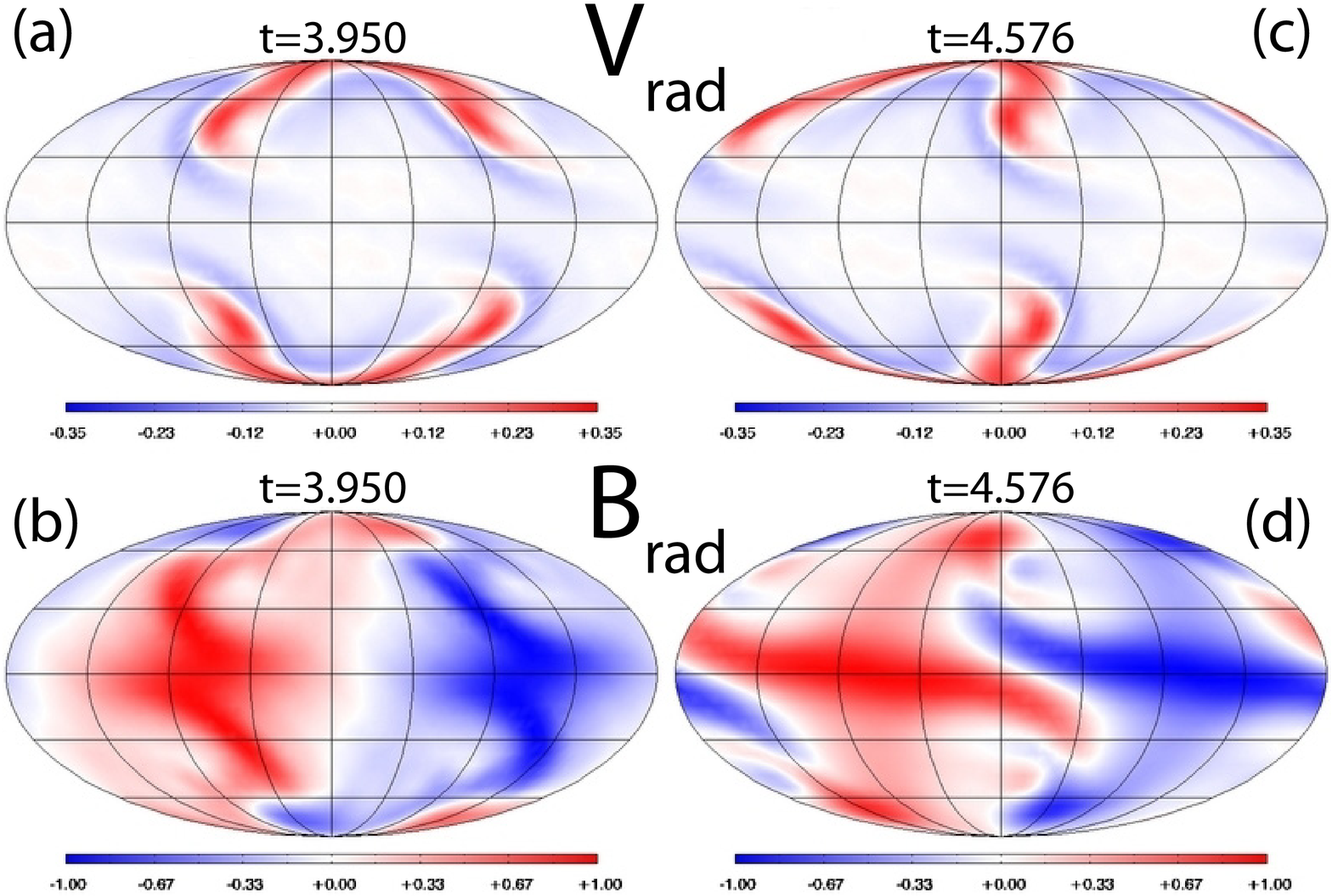}
\caption{\label{fig10}
Radial vector component of the velocity field (a, c) and the normalized magnetic
field (b, d) at 90\% of the sphere's radius before (a, b) and after (c, d) the
first transition from growth to decay, labeled NL.  The snapshots were taken at
maxima of the (2,0) kinetic energy time trace, viz.\ Fig.~\ref{fig08}~(b).
}
\end{figure}
\begin{figure}   \centering
\includegraphics[width=\columnwidth]{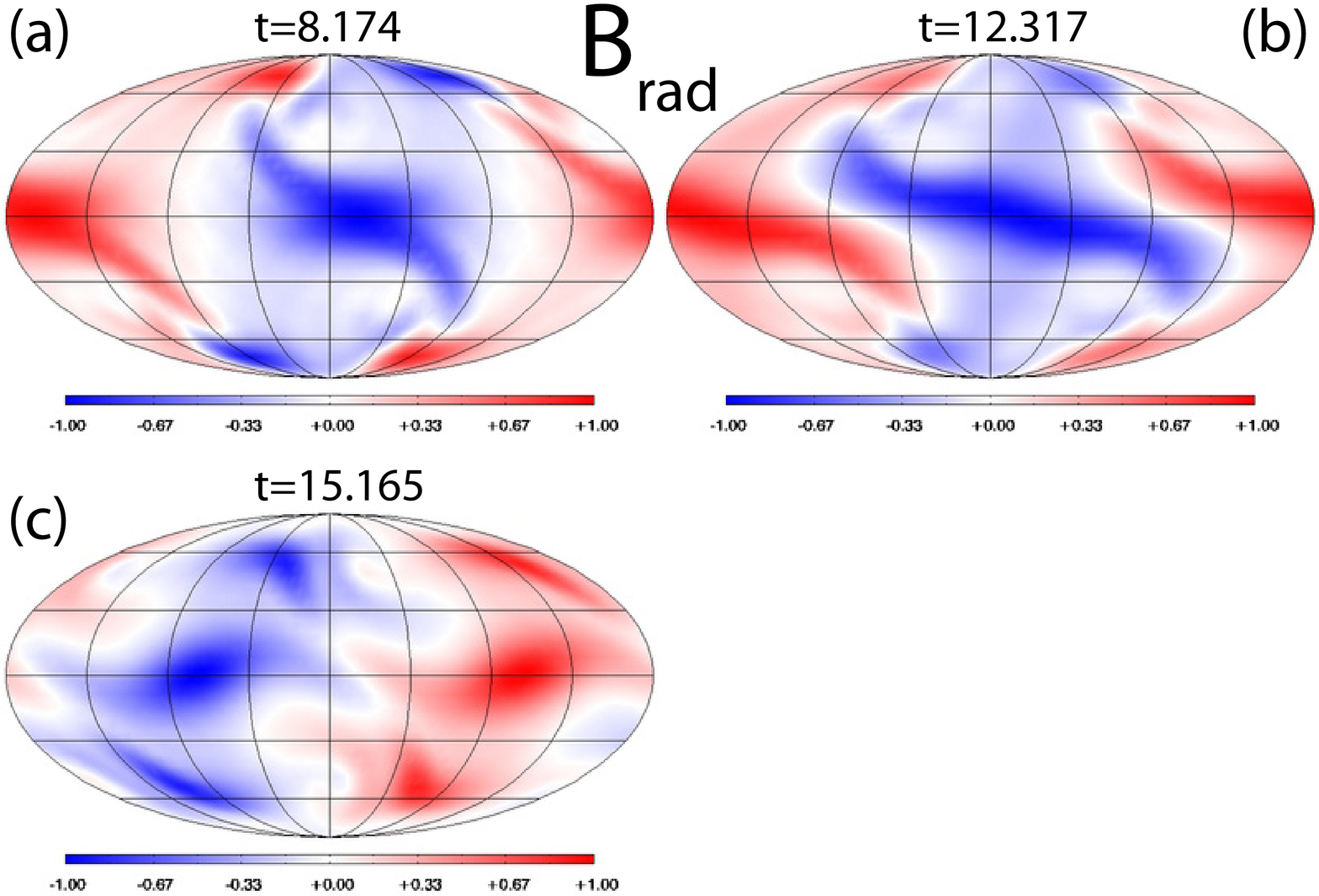}
\caption{\label{fig11}
Radial vector component of the normalized magnetic field at 90\% of the sphere's
radius during the second growth phase (a), the second decay phase (b), and the
third growth phase (c).
}
\end{figure}
To this end, we begin with an investigation of the velocity field during the
transition NL.  The panels (a) and (c) in Fig.~\ref{fig10} show snapshots of the
radial component of the velocity field taken at maxima of the (2,0) kinetic
energy time trace, before and after the transition, viz.\
Fig.~\ref{fig08}~(a).
For the following discussion, it is useful to introduce a virtual plane on a
circle of longitude which is defined by the z-axis and the longitudinal
coordinate at which the counter-propagating wave features in the upper and lower
hemispheres ``meet.''  The position of the plane varies between runs due to the
rotational symmetry the flow has about the z-axis before it becomes
hydrodynamically unstable.  However, once states like C or D are reached, this
plane stays fixed in space since the waves counter-propagate at the same
velocity in each hemisphere.  Relating this virtual plane to the oscillations in
Fig.~\ref{fig08}~(a), the wave features reach it at maxima of the (2,0) kinetic
energy time trace.
From Figs.~\ref{fig10}~(a) and (c) it is evident that the nonlinear feedback
during NL due to the Lorentz force rotates the plane by $\pi/2$ about the z-axis
which translates to the phase shift by $\pi$ relative to the unperturbed flow as
indicated in Fig.~\ref{fig08}~(a).
The panels (b) and (d) of Fig.~\ref{fig10} show the radial magnetic field
snapshots associated with the velocity fields shown in the panels (a) and (c).
The magnetic field is dominated by an (1,1) equatorial dipole mode which is
evident from the flux patches peaking on the equator.  The magnetic field's
geometry changes periodically in time during wave motion, however, the
equatorial dipole stays fixed sufficiently far from the transitions which can be
understood from the same symmetry argument as in the case of the aforementioned
plane.
The magnetic equatorial dipole does not change its position during NL as a
comparison of the panels (b) and (d) in Fig.~\ref{fig10} yields.
Comparing the panels (a) and (b) in Fig.~\ref{fig10}, the plane characterizing
the wave motion is co-aligned with the equatorial dipole before the transition.
After NL, the plane and the dipole are oriented perpendicular to each other due
to the changes in the velocity field, viz.\ Fig.~\ref{fig10}~(c) and (d).
Obviously, the alignment of the velocity field relative to the magnetic field is
decisive if the magnetic field grows or decays in time.
Let us continue with following the changes to the fields during subsequent
transitions.
Fig.~\ref{fig11} shows snapshots of the normalized radial magnetic fields during
the second growth phase, the second decay phase, and the third growth phase.
During the transition from decay to growth at $\tau_{\sigma} \approx 6.7$, the
equatorial magnetic dipole rotates by $\pi/2$ as it is evident from comparing
Figs.~\ref{fig10}~(d) and \ref{fig11}~(a), whereas the flow remains in the state
indicated by Figs.~\ref{fig10}~(c) and \ref{fig09}~(a).
Hence, the magnetic dipole component and the plane characterizing the wave
motion are parallel to each other again which results in magnetic field growth.
Once the magnetic field has grown sufficiently strong which is the case at
$\tau_{\sigma} \approx 10.1$ the wave motion is pushed towards the initial
state as depicted in Fig.~\ref{fig10}~(a) while the equatorial dipole keeps its
position, cf.\ Fig.~\ref{fig11}~(b).
The magnetic field decays and starts growing again when its equatorial dipole
component rotates by $\pi/2$ to reach co-alignment with the plane characterizing
the wave motion, cf. Figs.~\ref{fig10}~(a) and \ref{fig11}~(c).
In summary, the equatorial dipole tilts by $\pi/2$ about the z-axis between
subsequent growth phases which adds up to a field reversal after two cycles,
cf.\ Figs.~\ref{fig10}~(b) and \ref{fig11}~(c).
The magnetic induction Eq.~\ref{ind_eq} allows for such field reversals since if
$\mathbf{B}$ is a solution for a given flow, $-\mathbf{B}$ is a valid solution
as well.
Whenever the magnetic field reacts back on the velocity field, the flow is
forced to switch between the configurations shown in Figs.~\ref{fig10}~(a) and
(c).
The dynamo based on flow C operates similarly to flow D during the growth phase
as it is shown in Figs.~\ref{fig10}~(a) and (b), however, does not show cyclic
behaviour.

The power spectrum of the growing magnetic field in system D is given in
Fig.~\ref{fig12}.  The power spectrum of the decaying magnetic field is
virtually undistinguishable and was therefore not included.  Compared to C, the
odd $m$ modes are more pronounced which is due to the preference for even $m$
modes in flow D.
\begin{figure}   \centering
\includegraphics[width=\columnwidth]{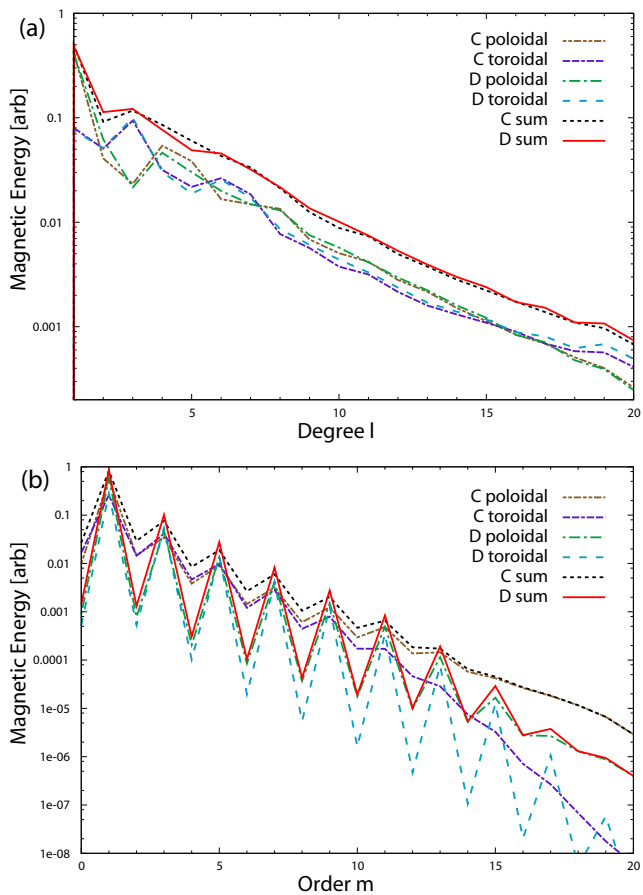}
\caption{\label{fig12}
Normalized and time-averaged spectra of the magnetic energy density during the
kinematic (growth) phase in the flows C and D as functions of spherical harmonic
degree $\ell$ (a) and order $m$ (b).
}
\end{figure}

\begin{figure}   \centering
\includegraphics[width=\columnwidth]{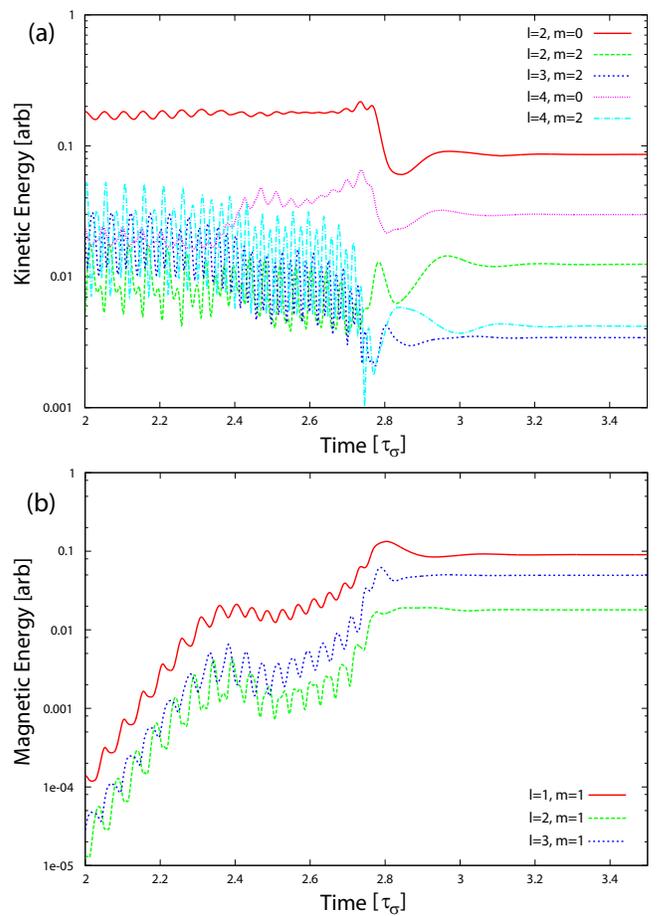}
\caption{\label{fig13}
Saturation of a dynamo based on flow D at $\mathrm{Rm}=51$.   The system evolves
towards a completely stationary state as it is illustrated by time traces of the
dominant kinetic (a) and magnetic (b) modes.
}
\end{figure}

Nonlinear saturation to a stationary state with $E_{mag}/E_{kin} \approx 1.2$ is
reached above $\mathrm{Rm}=51$.  Time traces of the dominant kinetic energy
modes during saturation are shown in Fig.~\ref{fig13}.  Compared to the
previously discussed cyclic dynamo the system saturates towards a completely
stationary state without any oscillations similar to the saturated dynamo state
of flow C at $\mathrm{Rm}=44$.
We now turn to a discussion of the physics results.

\section{Discussion}
\label{sec:conclusion}

Our numerical investigations indicate that magnetic field amplification
generated by time periodic non-axisymmetric $m=2$ hydrodynamic waves is relevant
to the large-scale dynamo process inside a spherical fluid conductor.
These waves emerge due to a hydrodynamic instability which breaks the rotational
symmetry of the flow.  The flow itself is driven by an axisymmetric body force
and subject to the zero-slip outer boundary condition in order to mimic a
situation similar to the Madison Dynamo Experiment.  Several bifurcation
sequences towards different statistically stationary flows with $m=2$ waves are
observed depending on the fluid Reynolds number.
The numerical investigations are performed in the intermediate regime between
the laminar flow and developed turbulence which facilitates the identification
of the dominant physical effects.

Let us briefly reiterate the key results which were obtained by solving the
induction equation using the aforementioned sub-turbulent flows.
At a given magnetic Reynolds number, the time-averaged flows as well as
snapshots of the flows are not dynamos whereas the time-dependent flows with
wave motion do show dynamo action.
By variation of the wave frequency, maxima in the efficiency of the dynamo
process are identified, as well as the breakdown of the dynamo for frequencies
which are too low or too high.
Moreover, one of the several possible statistically stationary states of the
flow shows cyclic dynamo action.  We find that the transitions between magnetic
field growth and decay crucially depend on the phase angle between
characteristic oscillations in the velocity and magnetic energy modes which
translates to changes in the relative alignment of the fields in physical space.
Hence, system D is different from the self-killing dynamos which have been found
previously \cite{Brummell1998,Fuchs1999}.

These phenomena are symptomatic of dynamo action based on non-normal growth
\cite{Tilgner_PRL_2008}.
Non-normal growth refers to perpetual growth due to the mixing of non-orthogonal
eigenstates of the time-dependent linear operator $\mathcal{L}(t)$ of the
induction equation, which can formally be written as $\partial \mathbf{B} /
\partial t = \mathcal{L}(t) \mathbf{B}$.
If the time dependence in $\mathcal{L}(t)$ given by a non-stationary flow is
adequate, perpetual growth of magnetic field modes is possible even if the
magnetic field would decay in the absence of time dependence.
Such mixing of non-orthogonal eigenstates was shown to act as an efficient
driver for dynamo action in Ref.~\cite{Tilgner_PRL_2008}, thereby identifying a
novel dynamo mechanism different from Lagrangian chaos or the $\alpha$-effect.
Using an infinite 2D periodic drifting velocity field which is no dynamo without
drift at a given $\mathrm{Rm}$, it was demonstrated that the velocity field
supports magnetic field growth for drift velocities not being too fast or too
slow.
The analogy between our results and the results presented in
Ref.~\cite{Tilgner_PRL_2008} strongly supports the conjecture that non-normal
growth is responsible for the magnetic field generation in the $s2t2$ flow with
wave motion.
Interestingly, flow C exhibits a wave frequency close to a maximum in dynamo
efficiency, cf.\ Fig.~\ref{fig06}.  This happens by chance since the wave
feature is a hydrodynamic phenomenon and, hence, is independent of a weak seed
magnetic field.

The observed saturation of the $s2t2$ dynamo to a state without wave motion
confirms the relevance of the adjustment of the wave frequency during the
saturation process \cite{Tilgner_PRL_2008}, cf.\  Fig.~\ref{fig13}.

Increasing the Reynolds number above $125$ leads to a superposition and
interaction of the $m=2$ waves with structures at larger wavenumbers.
Turbulence develops which manifests itself in an increase of the critical
magnetic Reynolds number $\mathrm{Rm_c(Re)}$, as it was shown numerically for
$s2t2$ systems \cite{Bayliss2007, Gissinger_PRL_2008, Reuter_NJP_2008}.
This increase in the stability curve due to turbulence is a big challenge for
laboratory dynamo experiments which use simply connected flows
\cite{Nornberg2006a, Spence2006, Spence2007, Monchaux2007, Ravelet2008,
Stefani_ZAMM_2008}.
Liquid sodium which is preferred as a working fluid has a magnetic Prandtl
number $\mathrm{Pm}=\mathrm{Rm}/\mathrm{Re}$ of the order of $10^{-5}$
\cite{Nornberg2006b}.
Hence, in order to achieve magnetic Reynolds numbers sufficiently large for
self-excitation, the experimental flows are vigorously turbulent with Reynolds
numbers $\mathrm{Re} > 10^5$, inaccessible to any numerical simulation.
Therefore, only extrapolations of $\mathrm{Rm_c(Re)}$ can be done.
In numerical studies of dynamos with $\mathrm{Pm}<1$ conducted in infinite
periodic boxes, which are usually computationally less expensive than
simulations in spherical geometry but lack realistic boundary conditions, the
stability curve $\mathrm{Rm_c(Re)}$ was shown to exhibit a saturation to a
plateau after an initial increase for large-scale dynamos \cite{Ponty2005,
Minnini2006, Laval2006, Ponty_NJP_2007}, and similarly for the fundamentally
different small-scale dynamo driven by non-helical random forcing, cf.\
\cite{Schekochihin2007} and the references therein.
In spherically bounded large-scale dynamos like the system we are focusing on in
the present study, the question if and at which level the dynamo threshold
saturates to a plateau at large $\mathrm{Re}$ has yet to be investigated.
Our study shows that field generation due to wave motion is important in the
context of the dynamo instability.  Beneficial effects are shown to exist in the
sub-turbulent regime at low Reynolds numbers of about $100$.
The question most interesting to experimenters is, of course, if these results
can be transferred to laboratory dynamos.
Simulation results obtained for Reynolds numbers above $1000$ are encouraging
since we still find a pronounced peak at $m=2$ in the spectra of turbulent
velocity fields.  This indicates that remnants of the $m=2$ waves could be
present in the form of coherent structures in experimental flows and potentially
act as a driver for magnetic field growth.
A modification one might suggest in order to exploit the findings of this paper
to improve $s2t2$ setups is the installation of deflector baffles in front of
the impellers, designed to force flow components with $m=2$ symmetry on top of
the axisymmetric background flow.  These baffles could then be rotated about the
impeller axis at variable frequency which would allow to search for an optimum
in frequency similar to the study presented in Fig.~\ref{fig06}.

Growing and transiently decaying magnetic solutions are supported by the flow D.
The question if the magnetic field grows is related to an alignment problem of
the velocity and magnetic fields.  The configuration of $\mathbf{v}$ and
$\mathbf{B}$ depicted in Figs.~\ref{fig10}~(c) and (d) constitutes a poor state
vector of the system since subsequently applying the linear operator
$\mathcal{L}(t)$ of the induction equation on $\mathbf{B}$ in order to evolve
the system forward in time---which is done by the numerical code---causes the
magnetic field to decay.  Note that the time-dependence of $\mathcal{L}(t)$ is
given by the temporal evolution of the velocity field starting from the
configuration shown in Fig.~\ref{fig10}~(c).
The modification of the flow during the transition from growth to decay is
certainly caused by the Lorentz force, however, the reason for the rotation of
the magnetic dipole during the transition from decay to growth is not obvious.
Slow diffusion of the magnetic field into the preferred position which is then
amplified to finally take over is a possible explanation.

Let us finally discuss the relation of the cyclic dynamo D to the competing
axial and transverse dipole modes which are reported for an $s2t2$ flow in
\cite{Gissinger_PRL_2008}.
Ref.~\cite{Gissinger_PRL_2008} describes the generation of an axisymmetric
magnetic field component via non-axisymmetric turbulent fluctuations in the
presence of an axisymmetric $s2t2$ background flow.  It is shown that, first, an
equatorial dipole magnetic field is generated which creates via Lorentz braking
an $m=2$ velocity field component which in turn generates the $m=0$ magnetic
field.  In the following, the $m=0$ and $m=1$ magnetic modes compete via a
modulation of the amplitude of the $m=2$ velocity mode.
However, these competing modes are fundamentally different from the cyclic
dynamo D which is presented in the paper at hand.
First, the $m=2$ non-axisymmetric hydrodynamic waves in our model emerge
independently of a magnetic field.
Second, when a weak seed magnetic field is introduced, the flows with $m=2$ wave
motion generate a dominant $m=1$ transverse dipole component.
The axisymmetric components of the magnetic field are relatively weak as the
power spectrum in Fig.~\ref{fig12}~(b) indicates.
Third, there are no competing modes observed.  The relative alignment of the
magnetic and velocity fields is decisive if the magnetic field grows or decays.
In summary, our results are by no means contradictory to the results reported in
\cite{Gissinger_PRL_2008}.  Differences in the forcing scheme and in the
turbulence intensity exist which are crucial if an axisymmetric dipole is
generated.  This is pointed out in section \ref{sec:hydro}.

In conclusion, magnetic field generation due to wave motion was shown to exist
in a spherical $s2t2$ dynamo model.  The transfer of these results to
experimental flows is not straightforward due to turbulence, nevertheless wave
effects could play a role in laboratory dynamos and should be kept in mind when
modifying existing or designing new experiments.

KR would like to thank C.~C.~Finlay and E.~J.~Spence for helpful conversations.
The simulations were performed on the TOK and BOB clusters hosted at RZG.  The
visualizations in Fig.~\ref{fig02} were created using the VAPOR visualization \&
analysis platform \cite{VaporPaper}.

\bibliography{references}

\end{document}